# Aeromechanics of Hovering Flight in Perturbed Flows: Insights from Computational Models and Animal Experiments

Chao Zhang[1]
*Shenzhen Escope Technology Co. Ltd, Shenzhen, 518052, China*

Tyson L. Hedrick[2]
*Department of Biology, University of North Carolina at Chapel Hill, Chapel Hill, NC, 27599*

Rajat Mittal[3]
*Department of Mechanical Engineering, Johns Hopkins University, Baltimore, MD, 21218*

and

Yijin Mao[4]
*Shenzhen Escope Technology Co. Ltd, Shenzhen, 518052, China*

**Stability of flapping flight, a natural requirement for flying insects, is one of the major challenges for designing micro aerial vehicles (MAVs). To better understand how a flying insect could stabilize itself during hover, we have employed a fully coupled computational model, which combines the Navier-Strokes equations and the equations of motion in six degrees-of-freedom (NS6DOF) to model the hovering flight of a hawkmoth. These simulations are combined with high-speed videogrammetry experiments on live, untethered hawkmoths flying in quiescent and perturbed flows. The flight videos were used to identify a potential mechanism that could be used by the moth to stabilize its hovering flight; the effectiveness of this mechanism was investigated using CFD-based simulations and semi-analytic approximations.**

## Nomenclature

$u_i$     = velocity component
$x_i$     = coordinate component
$p$     = pressure
$v$     = kinematic viscosity
$M$     = Mass matrix
$\vec{V}$     = velocity vector
$\vec{F}$     = force vector
$t$     = time
$I_B$     = pitch-related moments of inertia
$\vec{\Omega}_B$     = pitch rate of the body
$\vec{M}_W$     = angular momentum of two wings in pitch

---

[1] CEO, Shenzhen Escope Technology Co. Ltd.
[2] Associate Professor.
[3] Professor and AIAA Associate Fellow (Corresponding Author)
[4] CTO, Shenzhen Escope Technology Co. Ltd.

| | | |
|---|---|---|
| $\vec{T}_{aero}^{B}$ | = | external aerodynamic pitch torques physically exerted on the body |
| $\vec{T}_{aero}^{W}$ | = | external aerodynamic pitch torques physically exerted on the wings |
| $\vec{T}_{W-B}$ | = | internal pitch torque exerted by the flight muscles |
| $M(t)$ | = | aerodynamic pitch torque |
| $F^{X}(t)$ | = | aerodynamic force in horizontal axis |
| $F^{Z}(t)$ | = | aerodynamic force in vertical axis |
| $x_{cop}(t)$ | = | horizontal coordinate of the center-of-pressure (CoP) |
| $z_{cop}(t)$ | = | vertical coordinate of the center-of-pressure (CoP) |
| $q$ | = | pitch velocity of the whole body |
| $\theta$ | = | pitch angle of the whole body |
| $\vec{T}_{C}$ | = | internal control torque |
| $G_{C}$ | = | output gain for the Coriolis force sensory system |
| $G_{V}$ | = | output gain for the visual sensory system |
| $t_{C}$ | = | latency time of the Coriolis force sensory system |
| $t_{V}$ | = | latency time of the visual sensory system |
| $T$ | = | period of flapping cycle |

## I. Introduction

IN addition to developing the aerodynamic forces necessary to support themselves in the air, flying animals and their human-designed analogs, micro-air vehicles (MAVs) also must maintain stability when navigating an unpredictable aerial environment. Stability in flapping and especially insect flight has been the subject of many experiments, simulations and theoretical analyses over the past decade [1, 2, 3, 4, 5, 6]. These have largely focused on pitch, since these different avenues of investigation all typically show that pitch is unstable, with coupled oscillations developing in pitch and longitudinal motion unless damped out by large aerodynamic drag [7] whereas roll and yaw may be damped through a variety of other effects [8, 9, 10].

As expected from prior results for pitch stability, a three degree of freedom (3DoF) and six degree of freedom (6DoF) linear time-invariant (LTI) stability analysis developed from earlier hawkmoth computational fluid dynamic (CFD) results [11] indicate that dynamic (oscillatory) instability in pitch is one of the dominant modes of instabilities for a hovering hawkmoth. Thus, pitch stability for hawkmoths and most flying insects must be maintained through some feedback control mechanism. This mechanism may operate via adjustments to the wing motions, and wing motions associated with pitch stability or pitch torque production have been revealed by animal experiments[12, 13, 14, 15], physical simulation [16, 17], and computational simulation [4, 11]. Control may also be implemented through redirection of aerodynamic forces via changes in body configuration as noted below.

**Abdominal Flexion and Stroke-Plane Adjustments as a Strategy for Flight Stabilization**

It has been widely observed that hawkmoths perform abdominal flexion movements in response to changes in body pitch [18] or to changes in visual stimulus consistent with changes in pitch [19]. Although these experiments were conducted on tethered hawkmoths and the extent to which these responses play a role in free hovering flight is not yet fully resolved, it has been suggested that this is a plausible feedback control based strategy that is employed by hawkmoths in free flight [15]. Furthermore, Dyhr [39] also showed that abdominal flexion can be used to redirect lift forces and this mechanism could be sufficient to stabilize the hovering flight of a hawkmoth [4]. Additional evidence of such a mechanism can be seen in recordings of freely hovering hawkmoths; Fig. 1 shows qualitative evidence of abdominal flexion in the free hovering flight of a hawkmoth. This simple recording shows that: 1) abdominal flexion follows an oscillatory mode (schematically marked by yellow lines) at a frequency less than the flapping frequency and 2) the stroke plane of the flapping wings (schematically marked by blue lines) stays nearly fixed with respect to the lab reference frame. This second effect, fixing the stroke plane angle to the lab reference frame, is also prominent in free flight perturbation responses of hawkmoths recorded as part of this study (see Results).

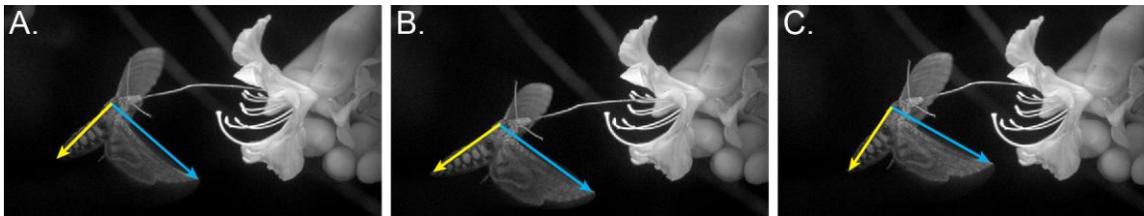

**Figure 1. Snapshots from high speed video (1000 Hz) of a freely hovering hawkmoth feeding from a still flower during training (a) 1st flapping cycle; (b) 5th cycle and (c) 8th cycle. The blue arrow connects the thorax to the wing tip and yellow arrow shows the location of the abdomen tip with respect to the thorax.**

**Sensing and Actuation for Hover Stabilization**

Feedback control of flapping flight requires sensing to detect deviations from the reference condition and actuation to generate forces and moments for control. The various sensory modalities of insect flyers have been studied extensively over the past decade [3, 21, 22, 23, 24, 19] and we employ some of this knowledge in developing a simplifed mathematical model for our pitch stabilizer. A widely available and well studied modality is the visual system. For instance, Fry et al. [3] conducted a series of experiments in a one-parameter open-loop paradigm to investigate the modulation of flight speed via the visual sensory system, and identified high-level control principles by applying genetic algorithms. Straw et al. [22] found the sensory-motor mechanism of motion control employed by Drosophila can be characterized as an integrated visual feedback approach in the horizontal and vertical planes simultaneously. However, visual sensory systems are slow and may impose a large latency between a perturbation and its response. Thus, fast mechanosensors such as the halteres of dipteran insects and antennal mechanosensors of hawkmoths [21] also serve a crucial role in the active control of insect flight. These mechanosensors may sense rotations through the Coriolis force and may also detect deviations in airflow. For example, the antenna of Drosophila, can also detect the wing induced airflow during vision-guided turns [24]. Thus, two sensing modalities, a long-latency sense for angular orientation and a short-latency sense for angular rate are incorporated into most models of insect flight control e.g. [14], including the model investigated here.

With regard to actuation, it is well established [25, 26] that many insects have two sets of muscles; one set of the muscles (usually indirectly attached to the wings) that power the flapping and another set (usually directly attached to the wing) used for fine control of the wing motion. This duality provides an effective paradigm for the actuation strategy required for flight stabilization.

**Prospectus**

This study uses quantitative recordings of freely hovering hawkmoths perturbed by vortices to identify a characteristic response of these animals to deviations in pitch orientation. This characteristic response - fixing the wing stroke plane to its lab frame orientation during steady hovering - are then examined in a CFD framework with 6DoF flow-induced motion (FIM). Effects of sensory latency on stability are examined, and the original animal recordings compared to FIM results of a simulated hawkmoth with closed loop control encountering a vortex ring perturbation. We find that, as expected, fixing the stroke plane does provide closed loop stability in the simulated moth even in the presence of realistic sensory latencies. The simulation results are also similar to the overall moth behavior in the perturbation experiments. Finally, stroke plane changes can be coupled with abdominal motion, essentially using the abdomen as the counterweight for overcoming wing inertia; this unites observed changes in abdominal pitch angle and wing stroke plane orientation. Thus, our results suggest a simple rule-of-thumb for maintaining pitch stability in flapping MAVs that also bears a striking resemblance to the behavior of flying hawkmoths.

## II. Materials and Methods

In this study we used a combination of live animal recordings from freely hovering hawkmoths, perturbation of these animals by directed vortex rings, Navier-Stokes computation modeling with flow-induced motion of the animal-vortex interactions, and a simple numerical model matching the output of the NS model. Details on each of these components are presented below.

**Animal Recordings and Vortex Perturbations**

We recorded and analyzed seven moth-vortex interactions in four male hawkmoths [Manduca sexta [27]] from the domestic colony maintained at the University of North Carolina at Chapel Hill. The moths were removed from the colony as pupae, allowed to eclose in mesh cages and subsequently maintained on a 20:4 h light:dark cycle to minimize activity in captivity and thus the accumulation of wing damage. Moths were provided with water ad libitum. Beginning three days post-eclosure, the moths were trained to hover and feed from an artificial flower by placing them in the flight chamber with dim visible-spectrum lighting near the end of their daily light cycle and introducing an artificial flower filled with 4:1 water:honey solution on a small support near the center of the chamber. Occasionally a live flower was used to elicit feeding interest and then replaced with an artificial flower. Once moths learned to visit the flower regularly they were considered candidates for experiments; moths were also maintained with daily feeding regardless of experimental activity. Recordings were collected when the moths were four to ten days post-eclosure.

Vortex rings for in-flight perturbation of the hovering moths were produced by a custom spring-loaded vortex cannon composed of a cylinder with a 66mm inner diameter, a plunger with variable plunge distance and a variable diameter exit aperature. For the moth experiments we used a 50mm aperature and plunge distances of 33 and 75mm. These produced rings which traveled at approximately $0.3$ ms$^{-1}$. The cylinder volume was seeded with smoke using a Safex fog generator prior to launch to allow visual tracking of the vortex and its interaction with the moths. During the recordings, a vortex was directed toward the moth in the horizontal plane along the anterior to posterior direction from a distance of approximately 10 cm as the moth hovered near the artificial flower.

The moth flights and vortex perturbations were performed in a 0.7x0.7x0.7 m glass-walled flight chamber with one face open to allow vortex targeting by the researchers. The chamber was dimly lit in the visible spectrum and brightly illuminated in the near-infrared (760 nm), below the moth's visual threshold, by eight infrared LEDs (Roithener LaserTechnik GmBH). Moth-vortex interactions were recorded at 1000 Hz with six high speed cameras arranged as three orthogonal pairs (two Phantom 7.1 and one Phantom 5.1, Vision Research and three Y4, IDT). The cameras were calibrated using a direct linear transformation for three-dimensional kinematic reconstruction [28]. Four points, the tip of the abdomen, center of the head and tip of the right and left wings were tracked manually, the average DLT triangulation residual was 0.55 pixels. These points were then used to compute the overall body pitch orientation and stroke plane angle of the moths before, during and after interaction with the vortex. Body pitch orientation was computed continuously as the angle between the vector running from the abdomen tip to center of head and the horizontal plane. Stroke plane angle was calculated for each half-stroke as the angle between the vector running from wingtip position at the start of the half-stroke to wingtip position at the end of the half-stroke and the horizontal plane and interpolated between the midpoints of the half-strokes using a cubic spline.

**Computational Modeling and Simulation Set-up**

The computational framework developed here involves flow modeling coupled with body dynamics modeling and it incorporates the ability to include models for sensing and actuation for flow stabilization.

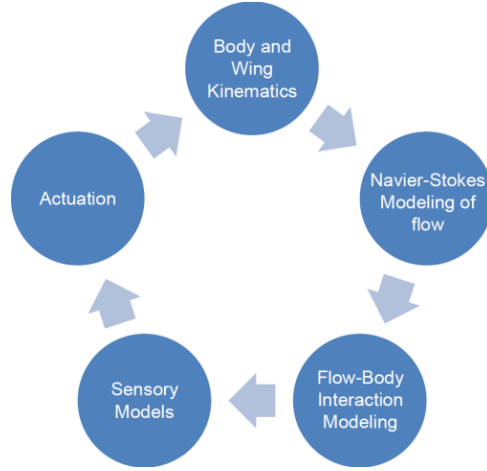

**Figure 2. Computational framework developed for exploring flight stabilization.**

In our study, a sharp interface immersed boundary method [29, 30, 31] is used to simulate the aerodynamics of the freely hovering hawkmoth [11]. The governing equations of aerodynamic flow during insect flight are the three-dimensional unsteady, viscous incompressible Navier-Stokes equations:

$$\frac{\partial u_i}{\partial x_i} = 0; \quad \frac{\partial u_i}{\partial t} + \frac{\partial u_i u_j}{\partial x_j} = -\frac{\partial p}{\partial x_i} + \nu \frac{\partial}{\partial x_j}\left(\frac{\partial u_i}{\partial x_j}\right); i,j = 1,2,3 \qquad (1)$$

where $u_i$ is the flow velocity component corresponding to direction i; $p$ is the flow pressure, and $\rho$ and $\nu$ are the fluid density and fluid kinematic viscosity, respectively. In this study, all the CFD simulations are conducted on a non-uniform 128×128×128 point Cartesian grid with 700 time-steps per flapping cycle. Detailed validation and grid convergence information for these simulations can be found in Zheng, et al. [32].

**Flow-Induced Body Dynamics (FIM) Simulation with Prescribed Wing Kinematics**

We employ a biologically derived wing shape and kinematics for our model; these are based on an earlier high-speed video recording and Navier-Stokes simulation of moth flight [11]. These produce the standard wing kinematics input for the simulation as in Fig. 3(b). The Reynolds number for the hovering hawkmoth is defined here as $\mathrm{Re} = U_{tip}\bar{c}/\nu$. Here $U_{tip}$ is the averaged wing tip velocity; $\bar{c}$ is the mean wing chord length; and $\nu$ is the kinematic viscosity. For this particular hovering hawkmoth, the mean chord length and mean tip velocity are measured to be $\bar{c} = 1.77 cm$ and $U_{tip} = 4.3 ms^{-1}$, respectively. The Reynolds number is specified to be $\mathrm{Re} = 1000$. The motion of hawkmoth is governed by a 6DoF equations of motion in the global frame fixed to the ground as shown in Eq. 2.

$$M\frac{d\vec{V}}{dt} = \vec{F} \qquad (2)$$

$$M = \begin{bmatrix} m & 0 & 0 & 0 & 0 & 0 \\ 0 & m & 0 & 0 & 0 & 0 \\ 0 & 0 & m & 0 & 0 & 0 \\ 0 & 0 & 0 & I_{xx} & 0 & 0 \\ 0 & 0 & 0 & 0 & I_{yy} & 0 \\ 0 & 0 & 0 & 0 & 0 & I_{zz} \end{bmatrix}, \vec{V} = \begin{bmatrix} u \\ v \\ w \\ \omega_x \\ \omega_y \\ \omega_z \end{bmatrix}, \vec{F} = \begin{bmatrix} F_x \\ F_y \\ F_z \\ M_x \\ M_y \\ M_z \end{bmatrix} \qquad (3)$$

where $\vec{V}$ and $\vec{F}$ denote the 6DoF velocity vector and force and moment vector respectively; $M$ is the mass matrix. However, this cannot be explicitly solved because the instantaneous force vector $\vec{F}$ is directly connected to the solution of Navier-Stokes equation in the following way:

$$\vec{F} = \int_B (p\hat{n} + \hat{\tau}) \cdot \vec{f} dS; \quad \vec{f} = \begin{bmatrix} \vec{e}_x \\ \vec{e}_y \\ \vec{e}_z \\ y\vec{e}_z - z\vec{e}_y \\ z\vec{e}_x - x\vec{e}_z \\ x\vec{e}_y - y\vec{e}_x \end{bmatrix} \quad (4)$$

where $\vec{e}_x$, $\vec{e}_y$ and $\vec{e}_z$ denote the unit vector in $x$, $y$ and $z$ axis respectively and the pressure $p$ and shear stress $\tau$ are calculated via solving the Navier-Stokes equation. The temporal discretization schemes of Eq. 3 are normally differentiated using an explicit or an implicit coupling scheme. The explicit coupling (also known as weakly or loosely coupled) scheme calculates the current time derivative by interpolation with the information of force and moment at previous time step as follows:

$$M^{n-1} \frac{\vec{x}^n - \vec{x}^{n-1}}{\Delta t} = F^{n-1} \quad (5)$$

A number of studies show that the explicit coupling scheme exhibits an inherent instability due to added mass effects when the density ratio (the ratio of solid density to the surrounding fluid density) is close to 1 or smaller [33, 34, 35, 36]. Many types of full or partitioned implicit coupling schemes have been developed as alternatives with good numerical stability, but at a much higher computational cost. However, because the density ratio of the hawkmoth examined here is very large ($\sim$ 800), explicit coupling is viable for this configuration and is the scheme of choice.

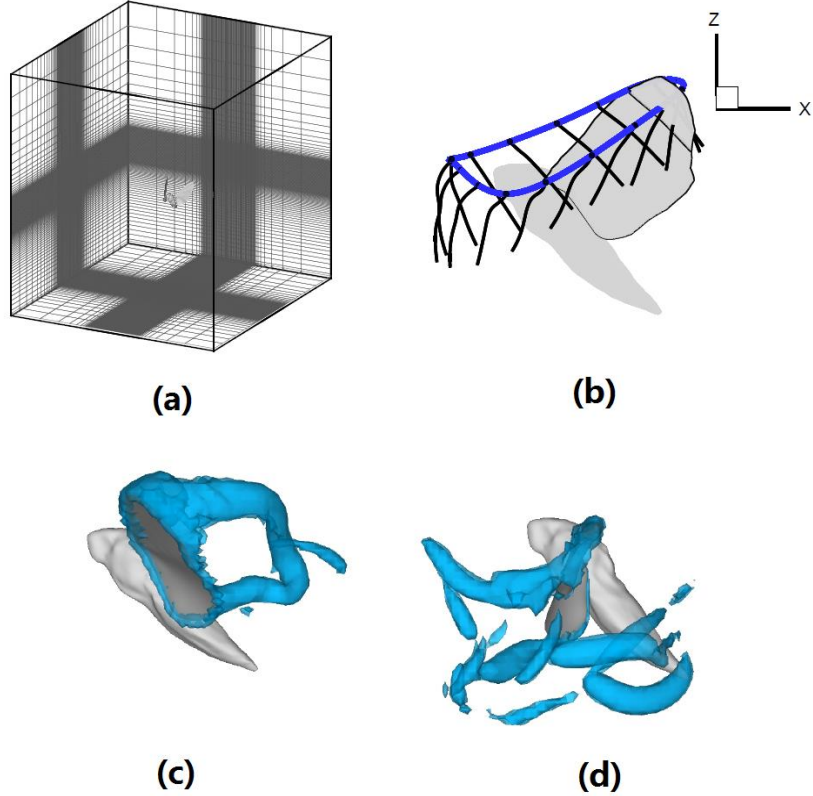

**Figure 3. (a) The mesh setup of hovering hawkmoth; (b) Standard wing kinematics throughout a whole flapping cycle featured by trajectory of 2/3 wing span leading edge point and the chordlines; (c) Vortex Structure of fixed moth with standard wing kinematics at t/T=0.31; (d) Vortex Structure of fixed moth with standard wing kinematics at t/T=0.77.**

**Mass Properties of Hawkmoth**

The mass properties of moth body and wings were measured from a moth population [11]. Although the mass of two wings accounts for only ∼ 0.07g (∼ 4% of the total mass), the contribution to the moment of inertia (MoI) is much greater (the averaged MoI component corresponding to pitch is up to ∼22% of the total MoI). Here we assume the mass distributions for the hawkmoth body and wings are uniform throughout, i.e. have constant density.

## III. Results

**Free-Flight Hawkmoth Vortex perturbations**

Figure 4 shows four video frames from a hovering hawkmoth encountering an incoming vortex ring. Rings moving in the horizontal plane from anterior to posterior along the moth typically pitched it upward and also pushed it backward. A principle wing motion response of the moth to these axial perturbations was to adjust the stroke plane angle of the wings with respect to the horizontal plane back toward the original, unperturbed position. Fig. 4 shows the complete time course of the moth's pitch perturbation in response to the vortex interaction along with the orientation of the wing stroke plane with respect to horizontal and also with respect to the moth's body axis. One wing beat after the perturbation arrives, the moth begins changing its wing orientation with respect to the body, reaching a maximum deviation near wing stroke 6. The overall effect of changing the stroke plane angle so as to keep it close to the original orientation rather than allowing it to move along with the body is a mechanism for stability that is explored further in this study via computational modeling.

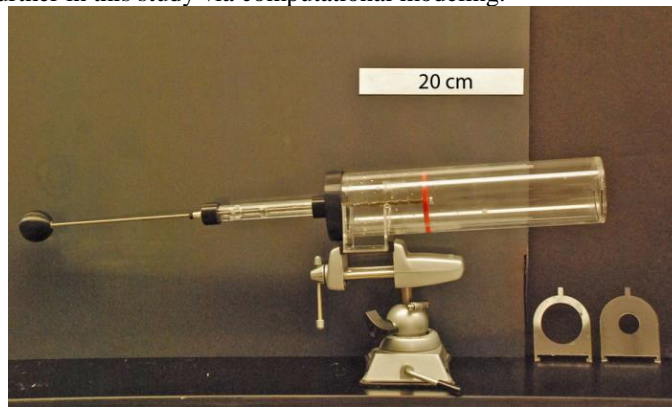

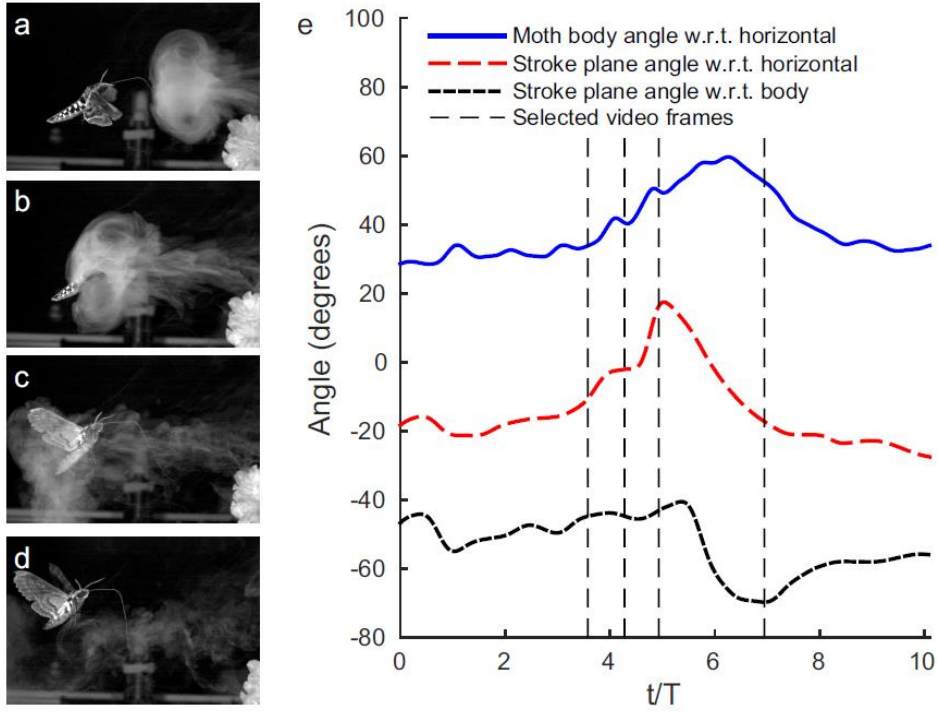

**Figure 4.** Top: Vortex cannon shown with the plunger fully extended and held in position by the stop ring; changing the stop position allows the plunger to be extended and then retract a fixed distance under spring tension. The different masks (lower right corner) allow production of a variety of vortex diameters. Panels a-d: hawkmoth experiencing a longitudinal vortex perturbation. Here the vortex reaches the moth at approximately mid-downstroke and results in a pitch-up perturbation of the moth. Panel e shows the time series of body pitch, wing stroke plane angle relative to horizontal and stroke plane angle relative to the body for the moth; dashed vertical lines correspond to the video frames shown in panels a-d. Note the decrease in the stroke plane angle relative to body during wing beats 6 and 7; this is part of the moth's flight control response to the perturbation.

**Free-Flight Hawkmoth Vortex perturbation simulations**

The first simulation conducted in this study was for a hawkmoth hovering in a quiescent flow with no feedback control implemented. In these simulations the Navier-Stokes are fully coupled to the 6DoF equations of motion for the hawkmoth body with wing kinematics fixed with respect to the body. The governing equations for this configuration are:

$$M \frac{d\vec{V}}{dt} = \vec{F}_{aero} - Mg\hat{k} \quad (6)$$

$$\frac{d}{dt}\left(\vec{M}_W\right) = \vec{T}_{aero}^W + \vec{T}_{W-B} \quad (7)$$

$$\frac{d}{dt}\left(I_B \vec{\Omega}_B\right) = \vec{T}_{aero}^B - \vec{T}_{W-B} \quad (8)$$

where $I_B$ is the pitch-related moments of inertia (MoI) of the moth body; $\vec{\Omega}_B$ denotes the pitch rate of the body and wings and $\vec{M}_W$ denotes the angular momentum of two wings in pitch; $\vec{T}_{aero}^B$ and $\vec{T}_{aero}^W$ are the external aerodynamic pitch torques physically exerted on the body and wings; $\vec{T}_{W-B}$ is the internal pitch torque exerted by the flight muscles to generate the flapping of the wings.

As expected, the uncontrolled or open-loop moth, when simulated using our FIM implementation is unstable in pitch, with noticable deviations for the initial state after a single wingbeat as shown in Fig. 5. The results demonstrate that the simulated hovering moth with kinematics and instantaneous wing shape matching an actual hawkmoth wingbeat is unstable in pitch.

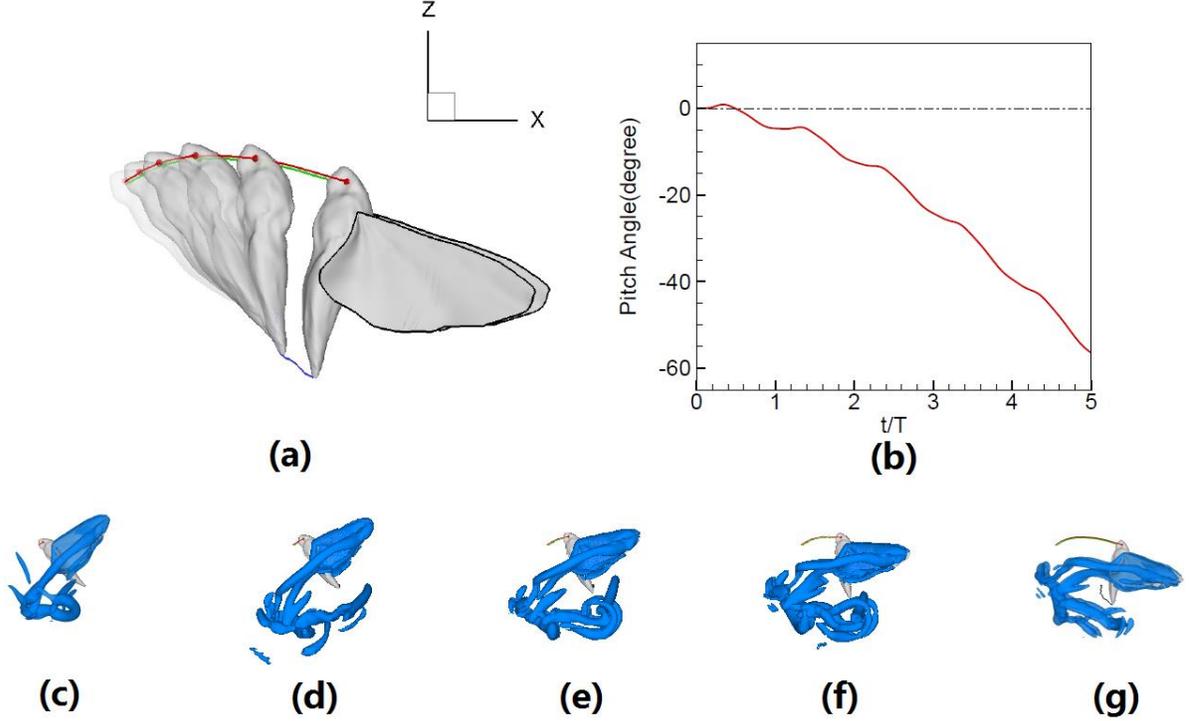

**Figure 5.** (a) The trajectory of motion of a simulated freely hovering hawkmoth throughout five flapping cycles without active control; (b) The pitch angle of the hawkmoth through time. A positive sign for pitch angle or pitch rate indicates anticlockwise rotation. The vorticity iso-surface of open-loop hovering moth is visualized at (c) t/T=1.03; (d) t/T=2.03; (e) t/T=3.03; (f) t/T=4.03; (g) t/T=5.03;

The instability is further demonstrated by analysis of the trajectory of the wing center-of-pressure (CoP) in time and the determination of neutral axes for pitch angle and pitch rate (see Fig. 6). The pitch neutral axis shows locations where, if the center of mass (CoM) was located on the axis, there would be no net change in pitch during a complete cycle; the pitch rate axis provides the same information for pitch rate. Thus, if the CoM were located at the intersection of the axes, the moth would be at equilibrium. However, the CoM is not located at this point. Furthemore, its position varies slightly over the flapping cycle and also changes over longer timescales due to natural activities such as feeding, dehydration and egg-laying. Furthermore, the neutral axes rotate with the moth, so deviations from equilibrium would not bring about their own restoring torque even if the CoM and neutral axes intersection were coincident. Due to the missing information about mass distribution of the moth body and inherent errors in the 3D segmentation, the location of CoM cannot be simply obtained by numerical calculation. However, given the wing kinematics, the force production at each time instant is known via numerical simulation. Fig 6(a) shows the numerically computed force vectors and the trajectory of the center of pressure in the longitudinal plane throughout the full cycle. If we consider that the center-of-mass of the moth body is fixed in space, then the pitch torque can be related to the force by the following equation,

$$M(t) = F^Z(t)\left(x_{cop}(t) - x_{com}\right) - F^X(t)\left(z_{cop}(t) - z_{com}\right) \tag{9}$$

where $M(t)$ is the aerodynamic pitch torque; $F^X(t)$ and $F^Z(t)$ are the aerodynamic force in horizontal and vertical axis, respectively; $x_{cop}(t)$ and $z_{cop}(t)$ are the coordinates of the center-of-pressure (CoP); In order to achieve an equilibrium state in pitch, we should also satisfy the following equation:

$$q(T) - q(0) = \int_0^T \frac{M(t)}{I(t)} dt = 0 \tag{10}$$

$$\theta(T)-\theta(0)=\int_0^T q(t)dt = 0 \qquad (11)$$

where $q$ and $\theta$ are the pitch velocity and pitch angle of the whole body. If we substitute Eq. 9 into Eq. 10, we can obtain the neutral line for pitch-rate (green line in Fig. 6(b)). The significance of this line is that if the CoM is located on this line, the net change of pitch rate over one cycle will be zero. Similarly, we can determine the neutral line for pitch-angle by solving the coupled solution of Eq. 10 & Eq. 11 and this is visualized as a cyan line in Fig. 6(b).

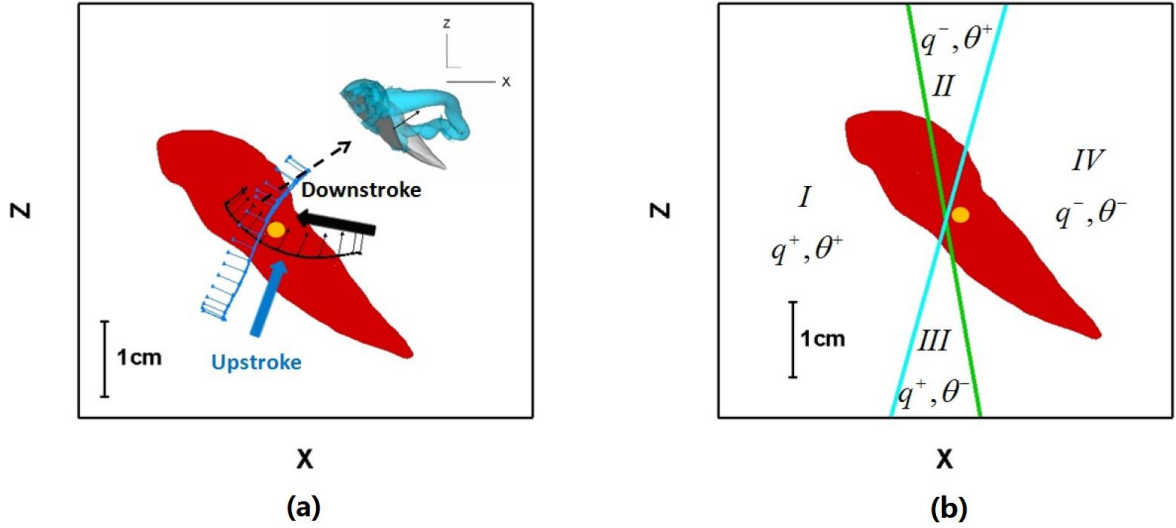

**Figure 6.** (a) Trajectory of instantaneous force throughout a cycle along with a sub-figure showing the instantaneous flow structure at t/T=0.4 (i.e. near mid-downstroke); (b) Neutral lines for pitch rate (Green) and pitch angle (Cyan). Positive sign of pitch angle or pitch rate indicates it is anticlockwise rotating and vice versa. "+" represents pitching down and "−" represents pitching up. Yellow dot is the COM.

In the live animal perturbation experiments, the moth appears to respond to the deviation in pitch by moving the stroke plane back toward its original orientation in the world frame. Applying an idealized version of this fixed-stroke plane response with zero latency and infinite gain does result in a stable pitch orientation in simulation. Figures 7(a), 7(b) and 7(h) show results of implementing this controller in our simulated hawkmoth with fully coupled 6DoF flow induced motion simulation. Figure 7(a) shows the trajectory of motion in six cycles with the trajectory of the eyes and abdomen tip color-coded so as to enable easy tracking of the motion of the moth. Figure 7(b) shows that by transferring the pitch torque from wings to body, the resulting relative pitch angle of hawkmoth body will cause a torque of the opposite sign on the body of the hawkmoth and the system eventually reaches a stable oscillatory state (i.e. limit-cycle) with an amplitude of oscillation of approximately 6 degrees. Furthermore, the final limit-cycle configuration has a mean relative pitch-angle between the body and the wing stroke plane of only about about 10 degrees different from the initial configuration, providing further evidence that the control strategy and assumptions about mass distribution within the moth are not unreasonable. If for instance, the final configuration had a much larger relative angle that would not be consistent with the experimentally based recording of the hawkmoth on which these simulations are based. Fig. 7(h) shows the movement of the CoM during the stable portion of the hover; we find that the stable configuration corresponds to an oscillation of the CoM about the neutral point that takes it back-and-forth between quadrants I and IV (as shown in Fig. 6(b)).

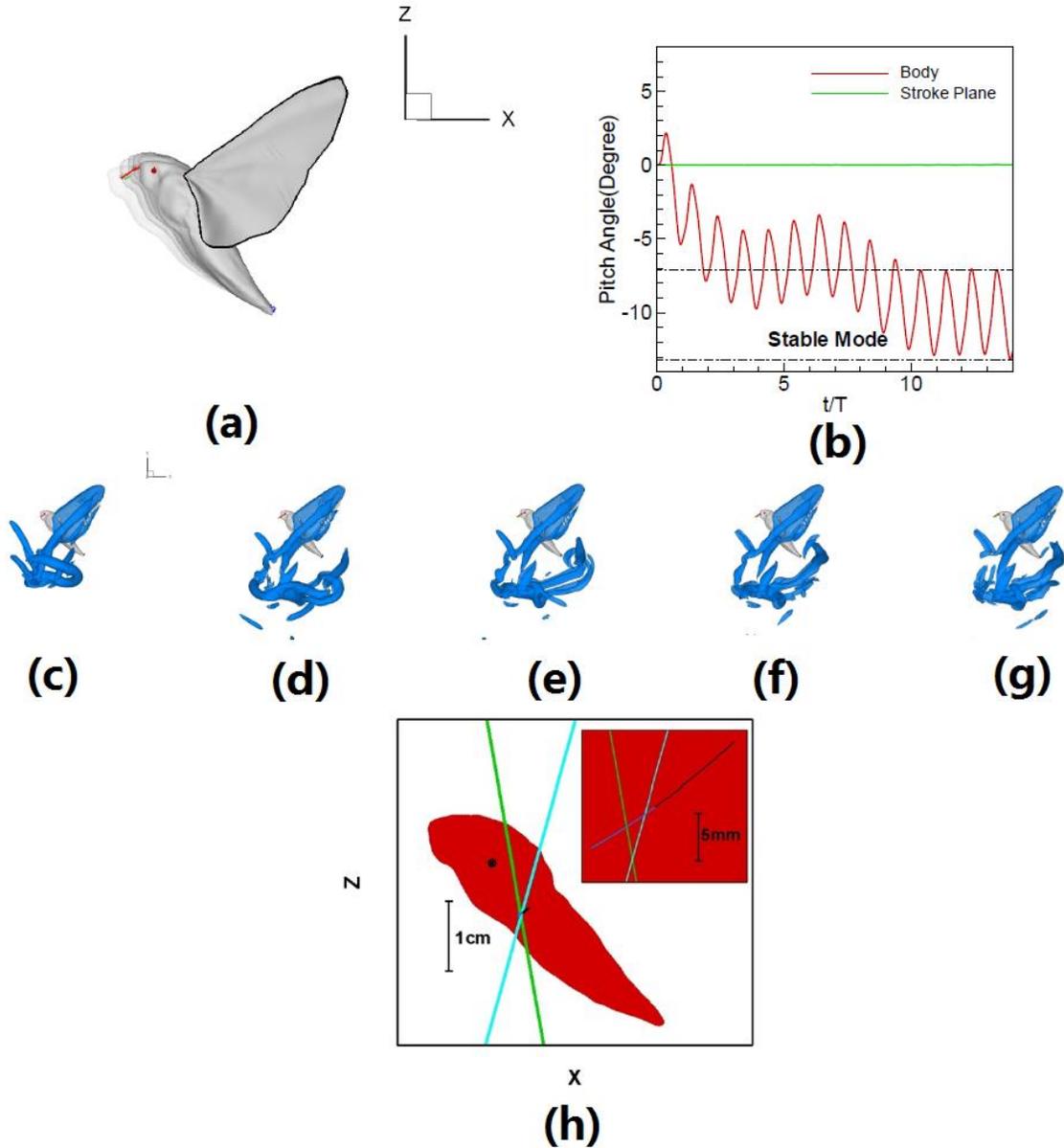

**Figure 7.** (a) The trajectory of motion of simulated free hovering hawkmoth throughout six cycles with the non-tilted stroke plane pitch controller in global frame; (b) The time series of pitch angle of the hawkmoth body and stroke plane (fixed); The vorticity iso-surface of hovering moth with idealized controller is visualized (c) at t/T=1.03; (d) at t/T=2.03; (e) at t/T=3.03; (f) at t/T=4.03; (g) at t/T=5.03; (h) The trajectory of COM in hinge-fixed frame.

## IV. Longitudinal Stabilization of Hover via Feedback Control

**Idealized Controller of Pitch Motion by Preventing Stroke Plane Tilt**

The change in angle between the body and stroke plane of the wing as well as the tendency of the hawkmoth to preserve the original inclination of the stroke plane (see earlier Results) inspired us to hypothesize that rotation between wing and body so as to to maintain the original, lab reference frame inclination of the stoke plane could be a possible strategy for hover stabilization. From the high-speed video, the stroke plane is observed to remain nearly constant throughout many cycles, despite changes in abdominal and body orientation. We also note that when the CoM is shifted to the left side of the neutral line of pitch angle (zone I and zone II in Fig. 6(a)), the net change of pitch angle θ will be positive (pitching down). However, if the position of the stroke plane is kept fixed, the increase of pitch angle will lead to the motion of the CoM back to the right side of angle neutral line, resulting in an upward pitching motion, and back and forth. This non-tilted stroke plane configuration coupled with body oscillation relative to the stroke plane could result in a stable loop because the instantaneous forces generated by the non-tilted stroke plane acts to damp pitch angle changes and prevent the CoM from going too far as it pitches up or down.

We tested this possibility by conducting a fully coupled 6DoF CFD simulation with an idealized controller embedded into the equations that couple the flow and the 6DoF dynamics model. This may be viewed as being governed by the following equations,

$$M \frac{d\vec{V}}{dt} = \vec{F}_{aero} - Mg\hat{k} \tag{12}$$

$$\frac{d}{dt}\left(I_W \omega \hat{j} + \vec{M}_W\right) = \vec{T}^W_{aero} + \vec{T}_{W-B} + \vec{T}_C \tag{13}$$

$$\frac{d}{dt}\left(I_B \vec{\Omega}_B\right) = \vec{T}^B_{aero} - \vec{T}_{W-B} - \vec{T}_C \tag{14}$$

where $\vec{T}_C$ is an internal control torque that is designed to impart an additional prescribed pitch-rate $\omega\hat{j}$ to the wings, transferring the angular momentum to the body. Here we assume zero latency in the sensory-actuation system and the control torque is chosen so as to cancel any deviation in the pitch of the wing from its equilibrium condition at each time-step.

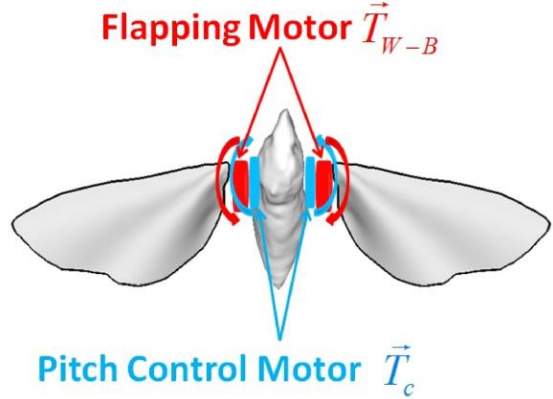

**Figure 8. Schematic of two sets of flight control actuators ('muscles' or 'motors') at the wing root that are used to modify the storke plane of the wings.**

**Neurosensory System Mediated Active Pitch Controller**

All sensory motor control systems in animals operate with some finite latencies and gains and the objective of the research described in this section is to examine the pitch-stabilization strategy with a more realistic model of the sensory motor control system of the hawkmoth. Thus, we incorporate latency and limits to gain in our controller, using a sensory system aware of pitch orientation via vision and pitch rate via Coriolis force sensing. The equation describing this type of sensory-motor control system for hover stabilization is written in terms of the pitch control torque as follows:

$$T_C = G_C \cdot \left(\omega^{W*} - \omega^W \Big|_{t-t_C}\right) + G_V \cdot \left(\theta^{W*} - \theta^W \Big|_{t-t_C}\right) \tag{15}$$

where $\omega^W$ and $\theta^W$ are pitch rate and pitch angle of the stroke plane (of two wings) and superscript "*" denotes the desired condition for the corresponding quantity; $G_C$ and $G_V$ are the output gains for the Coriolis force and visual sensory systems respectively, and $t_C$ and $t_V$ are the corresponding time-latencies. Note that the control torque is written in terms of the pitch rate and angle of the wings, this assumes that the moth is aware of the pitch orientation and rate of the body and also the position of the wings with respect to the body.

In order to start the exploration of latency and gain, we need some realistic estimates of the latencies and gains of these systems and for this we use experimentally determined values from Dickson et al.[14]. These indicate that the temporal response of the mechanosensory system is approximately one tenth of the flapping cycle while the visual system takes up to a full flapping cycle. We therefore prescribe the baseline latencies of the Coriolis force sensor and the visual sensor in the hawkmoth to be 0.1T and 1T respectively, where T is the period of the flapping cycle. There is very little understanding of what values of the gains are typical for these sensory modalities and we explored this in our simulation (see Results). The gains and latencies in both systems are normalized in the following way,

$$G_C^* = G_C / mgLT, \; G_V^* = G_V / mgL \qquad (16)$$

$$t_C^* = t_C / T, \; t_V^* = t_V / T \qquad (17)$$

where $G_C$ and $G_V$ are the gains for mechanosensory and visual systems, respectively; superscript "*" denotes a normalized parameter; $m$ is the total mass of hawkmoth (1.6g), $g$ is the gravity (9.8N/mg), $L$ is the span of a single wing (0.044m) and $T$ is the stroke period (0.04s).

**Linearized Model of Aerodynamic Torque**

Investigation of the effect of the control gains and latencies for the two sensory modalities on hover stabilization requires simulation of thousands of cases of flow-induced body dynamics. This is not feasible to accomplish with the full Navier-Stokes body dynamics coupled model because each of these simulations takes on the order of a week to complete on a large scale parallel computer cluster. In order to circumvent this problem, we have developed a linearized model of the aerodynamic torque based on quasi-steady aerodynamics as an alternative to the Navier-Stokes model. The model is based on the ideas of Dickson, et al. [14] who proposed this for investigating flight control models in the fruit-fly (*Drosophila melanogaster*). We focus here on the aerodynamic pitch torque, which is the key element in the stabilization and propose the following linearized model:

$$T_{aero} = T_{aero}^* + \left(\frac{\partial T_{aero}}{\partial \omega}\right)^* (\omega - \omega^*) + \left(\frac{\partial T_{aero}}{\partial u}\right)^* (u - u^*) + \left(\frac{\partial T_{aero}}{\partial w}\right)^* (w - w^*) \qquad (18)$$

where $T_{aero}$ is the total aerodynamic torque exerted on the hawkmoth and $\frac{\partial T_{aero}}{\partial \omega}$, $\frac{\partial T_{aero}}{\partial u}$ and $\frac{\partial T_{aero}}{\partial w}$ are the torque derivatives with respect to pitch rate, horizontal velocity and vertical velocity, respectively. For each of the variables $(u, w, \omega)$ we also compute two additional states by perturbing the particular variable. For instance, for variable $u$, solutions are also computed by imposing a perturbed forward flow velocity of $\delta u = \pm 0.01 m/s$ over the moth. Cycle averaged torque for these two conditions are computed and the derivative $\frac{\partial T_{aero}}{\partial u}$ estimated by employing a central-difference scheme as $\frac{\partial T_{aero}}{\partial u} = \frac{T_{aero}(u + \delta u, w, q) - T_{aero}(u - \delta u, w, q)}{2\delta u}$. The other derivatives $\frac{\partial T_{aero}}{\partial \omega}$ and $\frac{\partial T_{aero}}{\partial w}$ are also computed in a similar manner. For $w$, we assume perturbed values of $\delta w = \pm 0.01$ m/s and for $\omega$ the values employed are $\delta \omega = \pm 0.01$. Eq. 19 then replaces the Navier-Stokes equations in the parameter exploration of latencies and gains. Following the parameter exploration, we ran full Navier-Stokes 6DoF FIM simulations for selected cases.

**Effect of Feedback Gains on Stabilization**

The value of the gains $G_C$ and $G_V$ were varied from 0 to 0.29 and from 0 to 4.35, respectively in steps $\Delta G_C$ and $\Delta G_V$ of $2.9 \times 10^{-3}$ and $4.35 \times 10^{-2}$ respectively. This corresponds to a total of 10000 simulations, and this large search of the parameter space is made possible by the simple, linearized model of the aerodynamic torque. Each simulation is carried out for 20 flapping cycles and hover stability measured by the maximum cycle-averaged pitch (or tilt) angle of the stroke plane relative to its starting value. Fig. 9, shows this quantity plotted over non-dimensional parameter space of $G_C^*$ and $G_V^*$.

1. Except for two points where the stability boundary touches the x and y axis of the plot, non-zero gains for both the visual sensor and the Coriolis force sensor are required in order to stabilize the hover of the hawkmoth.

2. Due to the non-linearity of the dynamical system, there is a sharp bifurcation from stability to instability and this occurs once the pitch angle of stroke plane reaches 20 degrees.

3. The gains for the two sensory systems, that are most effective for stabilization and which confine the tilting of the stroke plane to within a low (below 5 degrees), range of tile angles are in the vicinity of $G_C^* = 0.2$ and $G_V^* = 1$.

The heat map shows that given a fixed value of latency for each sensory system, in order to stabilize the hovering hawkmoth in pitch, the magnitude of gains should be limited to a reasonable range, which is from 0.02 to 0.21 for $G_C^*$ and 0.0 to 0.3 for $G_V^*$. Here we should note that for the given latencies, theoretically the hawkmoth can stabilize itself even without visual sensory system.

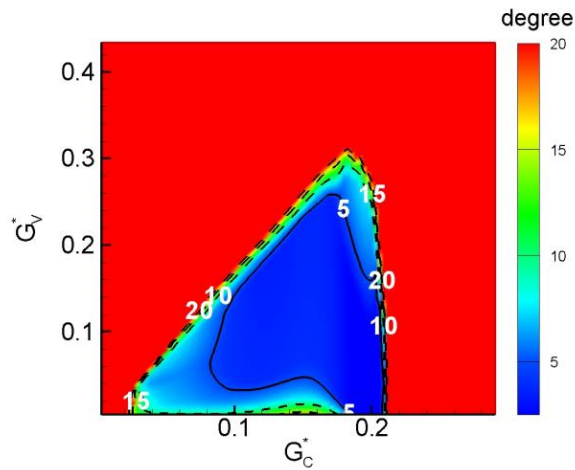

**Figure 9. Contour map of peak value of stroke plane tilt angle over 20 cycles correlated with the non-dimensional gains of mechanosensory (Coriolis force sensor) and visual feedback.**

In order to validate the results of the simple aerodynamic torque model used to generate Fig. 9, we conducted fully coupled CFD-6DoF simulations for two specific cases located in the stable region and unstable region respectively. In both cases, the non-dimensional latencies of the mechanosensory feedback and visual feedback are fixed as 0.1 and 1 respectively. In Case 1 (stable), we chose the non-dimensional gains of mechanosensory system and visual system to be 0.2 and 0.1 respectively while in case 2 (unstable), the value of the two non-dimensional gains are chosen to be 0.25 and 0.1. In case 1, the tilt angle of stroke plane is stably confined in the range of (−3°,3°) and the tilt angle of body is in the range of (−20°,5°) (see as shown in Fig. 10); while in case 2, a sudden surge of tilt angle occurs in 5 flapping cycles due to the over-amplified gains as shown in Fig. 11. Overall, the comparisons shown in Figs. 10 and 11 indicate that the linearized model provides a reasonably accurate representation on the initial dynamic of the moth dynamics.

Fig. 12 show the time variation of control torque and aerodynamic torque for these two cases and it is observed that in the stable case (see Fig. 12(a)) the control torque is one order-of-magnitude smaller than the aerodynamic torque; while in the unstable case, the control torque increases to very large values. For the stable case, the fact that the control torque is significantly smaller than the aerodynamic torque associated with the flappin is consistent with what is expected since the indirect muscles that power wing flapping are significantly more powerful than the direct muscles that generate small adjustments to the wing kinematics. From the viewpoint of bioinspired control also, this is a useful finding since is shows that a control system based on the current idea would need very small control inputs in order to stabilize a flapping wing vehicle.

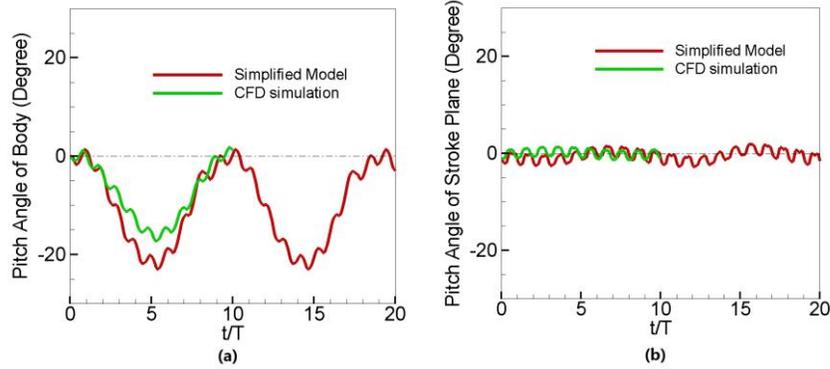

**Figure 10.** Time series of pitch angle of (a) hawkmoth body and (b) stroke plane obtained by simplified model (Red line) and CFD simulation and 6 DoF motion coupling under $G_C^* = 0.2$, $G_V^* = 0.1$, $t_C^* = 0.1$ and $t_V^* = 1.0$.

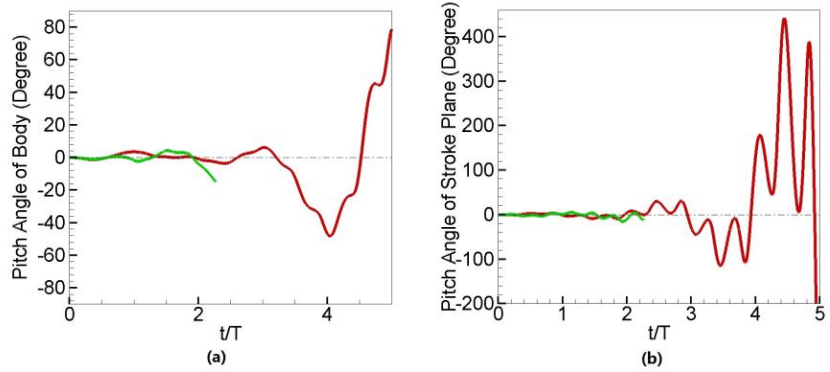

**Figure 11.** Time series of pitch angle of (a) hawkmoth body and (b) stroke plane obtained by simplified model (Red line) and CFD simulation and 6 DoF motion coupling under $G_C^* = 0.25$, $G_V^* = 0.1$, $t_C^* = 0.1$ and $t_V^* = 1.0$.

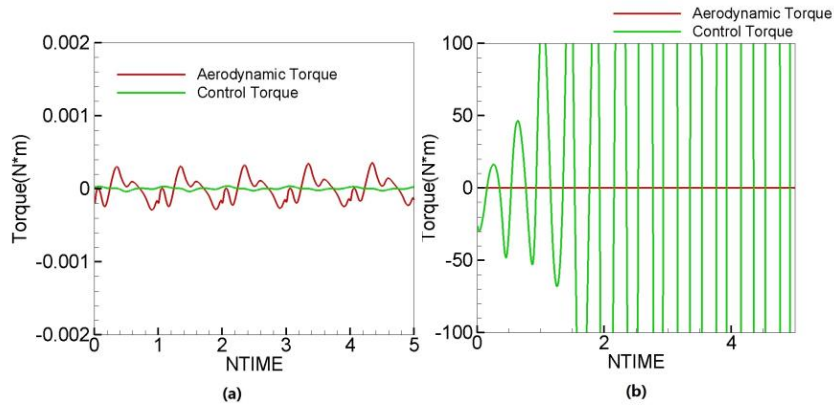

**Figure 12.** Time series of aerodynamic torque and control torque for the following cases: (a) $G_C^* = 0.2$, $G_V^* = 0.1$, $t_C^* = 0.1$ and $t_V^* = 1.0$ (b) $G_C^* = 0.25$, $G_V^* = 0.1$, $t_C^* = 0.1$ and $t_V^* = 1.0$

**Effect of Latency on the Sensory-Motor Control System on Hover Stabilization**

Using values for gain from the above investigation of varying gain and fixed latency, we explored the effect of latency on the control effectiveness. Values of $G_C^* = 0.2$ and $G_V^* = 0.1$ are chosen and we then varied the non-dimensional mechanosensory and visual latencies from 0 to 1 and 0 to 20 respectively, with corresponding step sizes of 0.01 and 0.2, respectively as shown in Fig. 13. As with our examination of gain, we found that the pitch stability is likely independent of visual sensory system latency; while it strongly relies on the latency of Coriolis force sensory system, which should be restricted to less than $0.1T$ to maintain its stability.

If we choose the combination of $G_C^* = 0.2$, $G_V^* = 0.1$, $t_C^* = 0.1$ and $t_V^* = 0.1$ to be the standard value for the gains and latencies, which is located well within the stable region of the map, then the time series of stroke plane tilt angle and moth body oscillation can be also computed via the simplified model, and these are plotted in Fig. 10. Note that not only is the stroke plane angle strictly bounded in the low range of tilt angle, the oscillations in the position of the hawkmoth body are also limited to a small range. Furthermore, a low dominant low-frequency oscillation is very apparent in the pitch angle of the body which is further examined below.

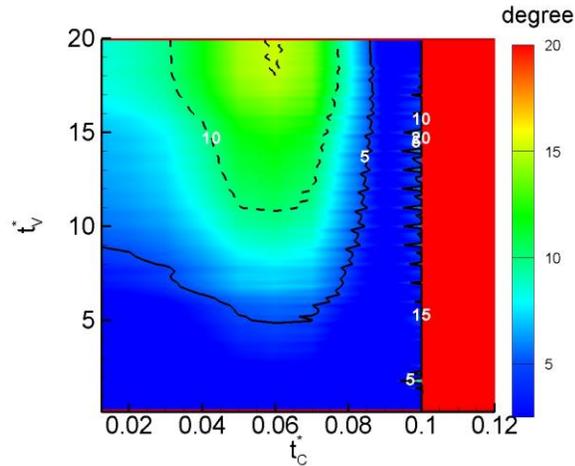

**Figure 13. Contour map of peak value of stroke plane tilt angle over 20 cycles correlated with the non-dimensional latencies of mechanosensory (Coriolis force sensor) and visual feedback.**

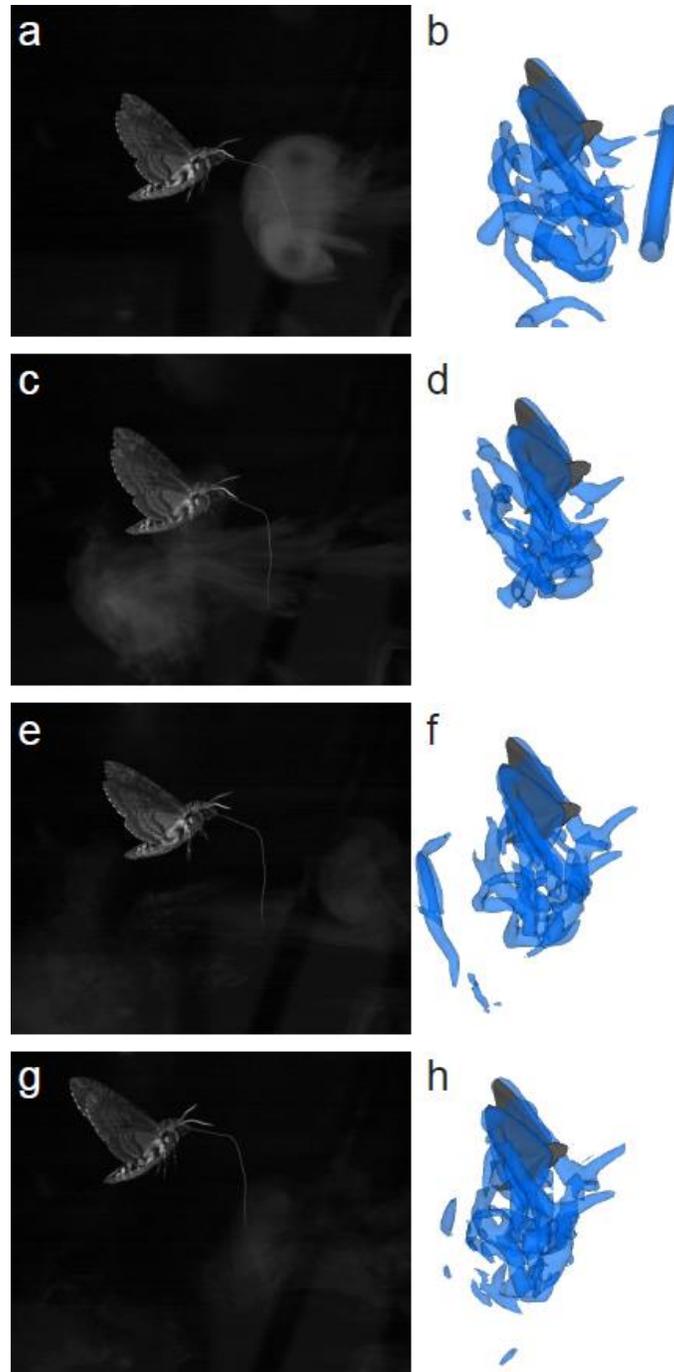

**Figure 14. Snapshots of free hovering hawkmoth from high-speed video (panels a, c, e, g) and from numerical simulation (panels b, d, f, h) at the 2nd, 4th, 6th and 8th cycles respectively.**

**Simulating Response of a Hovering Moth in Vortex Ring Encounters**

Here we examine the possibility that the the overall strategy of transferring pitch torque from wings to body for pitch stabilization via a constant stroke plane angle can be used for stabilization in the presence of large aerodynamic perturbations such as those encountered by natural and engineered microflyers. In the current study, we use the impact of a vortex ring to generate these large perturbation. In the numerical simulation, the vortex wing was generated by a jet impulse from the boundary of the computational domain with the same circular diameter and at the same distance as the actual vortex gun in the experiments. Here we used the idealized pitch controller with zero latency in conjunction with our fully coupled FIM 3DoF body-dynamics simulation. Fig. 14 shows a comparison of the experiment (left panel) and the simulation (right panel). The simulations are carried over the entire duration of the ring impact and its eventual passage past the hawkmoth, a total of about 8 flapping cycles and it is clear from Fig. 16 that the simulated moth maintains stability despite the large aerodynamic perturbation.

Figure 15 shows the time course of the pitch angle of the body and the wing with respect to their original values over the duration of the eight wing-strokes simulated here; we see that the initial pitch-up motion generated due to the impact of the vortex ring is damped out over 5 cycles (3rd to 8th cycle) by rotation of the body relative to the stroke plane. After the effect of perturbation is gone, the moth body would finally return to the oscillatory stable mode as shown in Fig. 7.

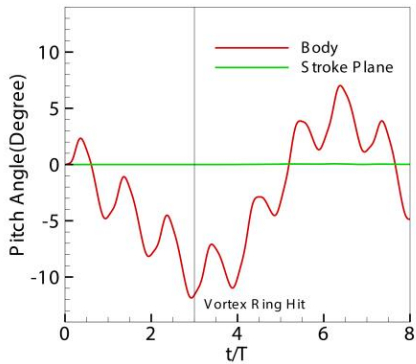

**Figure 15. Time series of pitch angle of body and stroke plane following vortex ring perturbation from the computational model.**

## V. Summary

We used high speed video recordings of hawkmoths experiencing in-flight aerodynamic perturbations to inspire a feedback control strategy for pitch-stabilizing hovering flight in a hawkmoth. We then tested this control strategy using 3DoF and 6DoF Navier Stokes simulations with flow induced motion, and futher developed simplifed LTI models to explore the parameter space of gains and latencies. The pitch controller functions by attempting to maintain the stroke plane of wings in its original orientation with respect to the laboratory frame, regardless of body orientation. This is implemented in our model by a pitch control actuator that introduces relative pitch motion between the body and the wings. Thus, body pitching motion is produced opposite the wing pitching motion required to keep the wing stroke plane steady in the laboratory frame. This mechanism differs from many previously hypothesized uses of the abdomen for flight stability [40, 39], where the envisioned mechanistic link is the change in position of the center of mass with respect to the center of pressure and is similar to [38] in that the angular momentum of the abdomen is used to affect flight force directions. Here we both extend and simplify this approach, linking the abdomen directly to the kinematics, enforcing a simple kinematic control rule which provides pendulum-like stability.

This approach directly addresses one of the common questions raised regarding insect flight stability - why do insects not have pendular type stability? Typically, linear-time invariant and other models of insect flight do not reveal pendular stability because, unlike a pendulum, the orientation of the forces supporting the moth are fixed to the body; if the moth rotates then the flight forces rotate also and there is no restoring torque generated. A pendulum generates forces fixed to the laboratory frame and therefore can provide a restoring torque. The control strategy of keeping the wing stroke plane fixed in the laboratory frame essentially recreates a pendulum. Because the wings and aerodynamic forces remain mostly fixed in the lab frame, restoring torques are generated by motion of the center of mass of the animal.

We took a number of additional steps to examine whether this stabilization strategy is reasonable for insects and MAVs. We implemented a simple feedback control mechanism parameterized by four parameters: two latencies and two gains. A simplified model of the aerodynamic pitch torque was developed based on the Navier-Stokes simulations and coupled with the 3DoF body dynamics equations to explore this large four parameter space; these simulations indicate that a mechanosensory latency of about $0.1T$ and a visual latency of $1T$ is ideal for hover stabilization. These values are in line with those for other insects [21,37]. The simulations also allow us to determine values of gains that provide the highest degree of stability and also the overall sensitivity of the control scheme to these parameters. A fully coupled Navier-Stokes-6DoF body dynamics simulation was conducted for two selected cases (one stable and one unstable) and the results are found to be in good agreement with the predictions from the simplified model. Comparisons from the simulation data are also made with an actual hovering hawkmoth and it found that the body pitch varies over a longer time scales (6 to 10 flapping stroke) in a way similar to that observed in the experiments. The simulations also indicate that with these values of gains and latencies, the sensory-motor-control system hypothesized here requires control inputs that are two-orders of magnitude smaller than that required to power the flapping of the wings.

The conclusion therefore is that the strategy of maintaining the stroke plane in its original orientation by allowing relative rotational motion between the wing and the body is effective for hover stabilization and might indeed be one strategy that is used by hovering hawkmoths. Such a strategy could also be used for small flapping wing MAVs.

## Acknowledgments

Funding from the following sources is acknowledged for this grant: Air Force Office of Scientific Research FA9550-10-1-006 (RM & TH); National Science Foundation CBET-1357819 and PLR-1246317 (RM); and National Science Foundation IOS-0920358 & IOS-1253276 (TH).